\documentclass[conference]{IEEEtran}
\IEEEoverridecommandlockouts
\usepackage{cite}
\usepackage{amsmath,amssymb,amsfonts}
\DeclareMathOperator*{\argmax}{argmax} 
\usepackage[]{algorithm2e}

\usepackage{graphicx}
\usepackage{subcaption}
\usepackage{textcomp}
\usepackage{xcolor}
\def\BibTeX{{\rm B\kern-.05em{\sc i\kern-.025em b}\kern-.08em
    T\kern-.1667em\lower.7ex\hbox{E}\kern-.125emX}}
\begin{document}

\title{Automatic Device Selection and Access Policy Generation based on User Preference for IoT Activity Workflow}

\author{\IEEEauthorblockN{1\textsuperscript{st} Mohammed Alshaboti}
\IEEEauthorblockA{\textit{Engineering and Computer Science} \\
\textit{Victoria University of Wellington}\\
Wellington, New Zealand \\
alshaboti.it@gmail.com}
\and
\IEEEauthorblockN{2\textsuperscript{nd} Aaron Chen}
\IEEEauthorblockA{\textit{Engineering and Computer Science} \\
\textit{Victoria University of Wellington}\\
Wellington, New Zealand \\
Aaron.Chen@ecs.vuw.ac.nz}
\and
\IEEEauthorblockN{3\textsuperscript{rd} Ian Welch}
\IEEEauthorblockA{\textit{Engineering and Computer Science} \\
\textit{Victoria University of Wellington}\\
Wellington, New Zealand \\
Ian.Welch@ecs.vuw.ac.nz}
}

\maketitle

\begin{abstract}
The emerging Internet of Things (IoT) has lead to a dramatic increase in type, quantity, and the number of functions that can be offered in a smart environment. Future smart environments will be even richer in terms of number of devices and functionality provided by them. This poses two major challenges a) an end user has to search through a vast number of IoT devices to identify the suitable devices that satisfy their preferences, and b) it is extremely difficult for users to manually define fine-grained security policies to support workflows involving multiple functions. 

This paper introduces an intelligent new approach to overcome these challenges by a) enabling users to describe their required functionalities in the form of activity workflow, b) automatically selecting a group of devices to satisfy users functional requirements and maximise their preferences over device usage, c) systematically generating local network access control policies to ensure enforce the principle of least privilege. We study different heuristic search algorithms to find the preferred devices for a given workflow. Experiments results show that the Genetic Algorithm is the best, among the algorithms that we test, as it offers a balance between efficiency and effectiveness.
\end{abstract}

\section{Introduction}
\label{sec:introduction}

The Internet of Things (IoT) paradigm is the next technological revolution in the era of computing \cite{madakam2015internet,furini2016iot}. Many of the objects that surround us are being replaced by networked things to form smart environments such as connected home, offices, buildings, and cities \cite{furini2016iot}. In 2018, the number of IoT devices used in smart homes and buildings has reached 1.2 billion, double any other IoT applications (e.g. smart cities or industrial IoT) \cite{statista2018IoT}. Tens or even hundreds of IoT devices are expected to be found in a shared smart environment such as a residential house or shared commercial premises in the near future. With the increasing number of consumer IoT devices, managing access control of these devices becomes very challenging.
This is compounded by the trend for automation to be no longer limited to the control of individual IoT devices (e.g. turning a single light on or off). Users can automate an activity that may involve multiple IoT devices. For example, a user can create an activity so that when s/he wakes up the alarm clock turns light on, opens the window, and triggers a coffee maker to start brewing. Such automation is now supported by many IoT automation frameworks such as   NodeRed \cite{nodered}, Stringify \cite{stringify}, Zapier \cite{zapier}, and Microsoft flow \cite{microsoftflow}. These frameworks utilise flow-based programming \cite{morrison2010flow} to facilitate automation of IoT-enabled activities. 

Though automating activities in smart environment may increase the quality of life, it is at the cost of the risk of compromised IoT devices. Which attackers can utilise to perform denial-of-service attacks, move laterally within the internal network or take over control of other IoT devices \cite{ding2018safety, yu2015handling}. For instance, in the previous example, a compromised alarm clock can open the window for a physical breach. Many solutions have been proposed for enhancing IoT security \cite{CorserIoTsec2017, zhang2010extended, su2018lightweight, yuan2005attributed,ye2014efficient}. Among them many are based on extending existing security methods (e.g IDS \cite{su2018lightweight}, IPS \cite{poseidon2017}, Network Isolation \cite{CorserIoTsec2017}) to suite new IoT scenarios. For example, intrusion detection requires establishing a baseline of normal activities performed frequently by users and detect unusual activities as abnormal events. Another approach is to enforce least privilege using network Access Control Lists (ACLs) which define what connections should be allowed or denied \cite{ouaddah2017access,amann2015providing, hamza2018combining}. ACLs policies to be defined manually by network administrators. We would not expect ordinary users to have the expertise to define such policies. Instead we propose an approach where users declare what functions they want to automate and automatically a) select suitable IoT devices based on users' preferences, b) generate a security policy that enforces least privilege among the selected devices.

In this paper we adopt the following threat model. There are different types of possible IoT attacks due to vulnerabilities in confidentiality, integrity, and access control. A systematic evaluation of privacy and security for consumer IoT devices, presented in \cite{loi2017systematically}, shows that the majority of threats are due to vulnerabilities in access control. Unauthorised access attack arises mainly from two sources \cite{hossain2015towards}: a) attacks from the local area networks (LANs) \cite{sivaraman2016smart}, and b) attacks from the Internet \cite{loi2017systematically}. Unauthorised access from the LAN can be originated from i) other IoT devices due to cross-device interactions, and ii) general purpose devices resides inside the local network. The latter can be mitigated by network segmentation \cite{CorserIoTsec2017} and is not the focus of this paper. This paper addresses local threats against IoT devices in the situation when IoT-to-IoT unauthorised access can happen in a smart environment such as fog computing in smart home networks  \cite{ding2018safety, yu2015handling}.


Existing works using offline/pre-defined policies such as  manufacturer usage descriptions (MUD) \cite{lear2018mud} are not sufficient for dynamic IoT environments \cite{hamza2018combining}. The MUDs are defined offline by IoT vendors to specifiy what network access is required for a particular IoT device to work properly \cite{lear2018mud}. For example, a coffee maker may only require inbound and outbound access to a cloud service on TCP port 443. MUDs can also be defined by a security expert \cite{ sorensen2017automatic}. However, in general predefined policies are not fine-grained enough, especially for local connections. For example, MUD may allow a device to establish local connections but it can't specify precisely the IP addresses and ports that devices should have access to. This is due to the context-dependent nature of device usage. For example, imagine a scenario where a user wants the morning alarm clock to trigger the coffee maker for the morning coffee. In this case, the least privilege would restrict network access so that only the alarm clock is allowed to access the coffee maker, rather than every single device in the network. You cannot specify this with a MUD, the best that a MUD can do is restrict access to being between devices by specific manufacturers. This doesn't take account the preferences or the needs of the user. Therefore, there is a need to generate automatically and dynamically ACL policies according to how devices are actually be used to support users' new and ongoing activities. 

One approach to automatic policy generation is to capture all kinds of benign traffic in relation to any targeted IoT devices offline \cite{hamza2018clear,sorensen2017automatic}, and parse the captured traffic to generate a policy to be enforced in the deployment \cite{ hamza2018clear,sorensen2017automatic}. Although this approach can generate fine-grained policies, it requires the training process to cover all possible usage of a device which is unfeasible for many IoT devices. Further, this approach doesn't follow the principle of least privilege because it difficult to determine precisely all allowed local endpoints since some aren't specified until deployment takes place. In view of this problem, we have developed a new technique for automatic policy generation for IoT devices based on the activities workflows provided by the users at the deployment.


We base our technique around the concept of workflow that doesn't specify which device to use. Workflows are common way to build and share automated activities in IoT automation platforms such as NodeRed \cite{nodered}, Stringify \cite{stringify}, and Microsoft flow \cite{microsoftflow}. Users are not always interested in manually specifying all the devices to be used for any activity. They are more interested in describing the required functionalities in an abstract way, for example, in the form of a flow of functions. However, currently a workflow is often tied to the underlying devices \cite{ giang2015developing, blackstock2016fred,szydlo2017flow, nodered, stringify}, such that a user can't use any pre-defined workflows unless he/she has exactly the same collection of devices specified in the flow \cite{nodered,stringify}. However, s/he may have alternative devices that can be used to accomplish the same activity.

\begin{figure}
    \centering
    \includegraphics[width=0.45\textwidth]{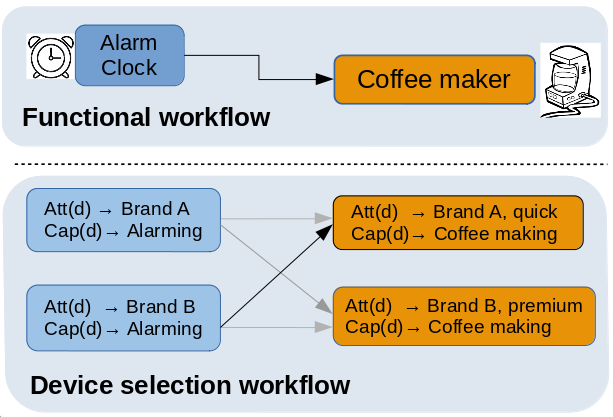}
    \caption{Functional workflow abstraction}
    \label{fig:wf}
\end{figure}

Decoupling workflow's functions from underlying devices enables users to share and re-use many existing flows. The concept of workflow abstraction based on device function is depicted in Fig. \ref{fig:wf}. For example, instead of building a flow by selecting the devices, Brand\_B alarm triggers Brand\_A coffee maker, flow can be defined in an abstract way as alarm triggers coffee maker, see \ref{fig:wf}. At the deployment our proposed system can automatically select the underlying suitable devices (e.g. Brand\_B alarm and Brand\_A coffee maker) based on the flow requirements and user preference. Hence, users can focus on what the activity function rather than the device selection  which can be tidies task in large IoT network. 

Further supported by a user preference model, our proposed approach can automatically select the suitable alarm and coffee maker to use from multiple alternative devices by maximising user preferences. In the context of IoT, user preferences might capture concerns related to security, privacy or even perceptions of quality. For example, users may prefer to use $Brand\_x$ devices for a sensitive function such as  audio recording due to the security reputation of their manufacturer. Other factors that may derive users' preference are how good or bad a manufacturer's privacy agreement and  the quality of the device, for example, the flavor of coffee made by coffee makers. Hence while selecting devices with respect to any workflow, users preferences must always be treated with priority.


Existing research has considered the problem regarding how to automatically predict users activities using a set of sensors \cite{rashidi2009keeping,wu2016bayesian}. The key idea is to analyse  user activities to predict what activities user may need next. This paper will propose a complementary approach, as we assume that the workflow for any important activity is given a priori, perhaps by using the methods developed in \cite{rashidi2009keeping,wu2016bayesian}. Our system subsequently finds the most preferable set of devices to satisfy users workflow by fulfilling all functionality requirements in the workflow and maximising users preferences. All kinds of information about a device can be used to build a preference model for a user. There are many excellent technologies to accurately learn preference models in the literature \cite{pigozzi2016preferences}. Any such technology can work well in our proposed system. Based on the chosen devices network access control policy can be established automatically to facilitate inter-communication among these devices while obeying the least privilege principle. 

Our main contributions of this paper are:
\begin{enumerate}
    \item Instead of binding workflow activities/applications to specific devices, we successfully decouple the functional flow of IoT applications from the underlying devices to enable easy sharing and flexible re-using of existing IoT workflows.
    \item We present the first mathematical formulation of user activities automation as a constraint optimisation problem with the goal of selecting a set of required devices to fulfil functional requirements and maximise users' preferences.
    \item We also develop a simple method to systematically generate fine-grained network access policies to support users' activities in a secure manner.
    
\end{enumerate} 

The rest of this paper is organised as follows. Section \ref{sec:relatedwork} presents related works. In Sections \ref{sec:netmodel} and \ref{sec:prefmodel}, we formulate network model and user preference model. Then in Section \ref{sec:problem_formulation} we present the formulation of the device selection problem. The algorithms studied by us to address device selection is further presented in Section \ref{sec:searchingmethods}. Based on the solution, the technique for automatic policy generation is subsequently detailed in Section \ref{sec:policygen}. Evaluation and results follow in section \ref{sec:eval} and conclusion in section \ref{sec:conclusion}.

\section{Related work}
\label{sec:relatedwork}
In this section we introduce the related research in access control for cross-device IoT interactions. And illustrate the related challenges, and highlight the technical and practical importance of automatic policy generation. We then discuss the related works in workflow, user preference, and least privilege access principle in the IoT domain. 

Researchers proposed different access control models such as Mandatory access control (MAC), discretionary access control (DAC), and role based access control (RBAC). MAC  requires subjects (e.g. users) to be classified into different security levels based on their role/trustworthiness. Objects will also be categorised similarly
based on their sensitivity \cite{sandhu1994access}. Access policy decides which subject class can access what object category. MAC is not entirely suitable for IoT. The interaction between the devices can't be easily mapped to classes, and IoT required a level of dynamicity which is missing in MAC \cite{guoping2011research, zhang2010extended, ouaddah2017access}. In contrast to MAC, DAC enables users to set access rules for the objects they own. Substantial research has been performed to include end-users input to build IoT security control \cite{neto2016aot,pediaditakis2012home,jindou2012access, liang2015sift}. For example, Neto et al. \cite{neto2016aot} proposed a cryptographic suite protocol (AoT) for authentication and access control which is driven by users input. In \cite{jindou2012access}, authors integrated RBAC with social network services and allowed end-users to define personalised access policy to govern device sharing. IoTSec \cite{CorserIoTsec2017} proposed a policy language that allows users to expressive policy for cross-device interactions. In these research works, users input drives fine-grained access control in consideration of social and environmental factors. However, heavily relying on users' involvement, increasing users' burden and prevent wide adoption. Therefore, this work proposed to automatically generate the policy for IoT given an activity workflow.

Workflow has been supported by many IoT frameworks such as NodeRed \cite{nodered}, Stringify \cite{stringify}, and Microsoft flow \cite{microsoftflow} to realise network activity automation. Researchers also used activity workflow to optimise latency and reduce bandwidth and utilise fog computing to enable distributed processing of activity workflow \cite{szydlo2017flow, giang2015developing}. Liang et al. utilised IoT data flow to ensure safety by verifying that any flow should never violate the safety policy \cite{liang2015sift}. We propose to use IoT activity workflows to a) automatically select the most preferable devices from a pool of candidates and b) generate network policy to enforce the least privilege access across the selected devices.

Researchers have studied the possibility of automatically deriving and predicting users activity \cite{rashidi2009keeping,wu2016bayesian}. Rashidi et al. \cite{rashidi2009keeping} proposed Apriori algorithm to detect activity patterns in smart home environment to automate users activities. Similarly \cite{wu2016bayesian} used Bayesian network to learn and predict users activities which can be used to generate personalised activities. These works focused predicting and automating users activities which can be represented as workflow, however, it didn't tackle how to build it and which devices will be used. Therefore, our goal is to a) select all devices that fulfil a workflow requirements and b) Choosing devices from the selected list that maximise user preference. For example, a user may prefer to use a brand $A$ devices for temperature sensing while prefer brand $B$ for thermostat function. In association with selected devices, policies will be generated automatically to ensure least privilege access, and segregate devices used by different users to protect their privacy. 

The main objective of MUD is to introduce the least privilege access required for IoT devices. However, as MUD policy is generated prior to the deployment it cannot define all access endpoints with fine granularity, especially local ones \cite{hamza2018combining}. In contrast to the MUD policy, we propose to use a user defined activity workflow to drive automatic policy establishment. Trimanada et al. \cite{trimananda2018vigilia} proposed Vigilia, which uses capability-based RMI access control to derive access control policy. However, it requires predefined capability-base access to specify the access between different type of devices prior to the depolyment. This requirement limits the interaction of a device to what is pre-defined in its driver. Sorensen et al. \cite{sorensen2017automatic} developed a firewall that  automatically generates profiles (i.e. policies) for IoT devices by monitoring their traffic during learning phase. Then enforce the learned policies during the depolyment. This approach work well, but it requires the devices to generate all possible traffic during the learning phase. However, we target the activity automation workflow so that policies can be automatically generated to ensure least privilege access to devices that are part of the workflow



\section{Network model}
\label{sec:netmodel}
We model a network as a collection of devices represented by  $N=\{d_i: i = 1,2,3,...,n\}$.
Each device $d$ in the network is represented by a tuple of a set of attributes $Att(d)$, a set capabilities $Cap(d)$, and a set of network requirements for a particular function $f$, given by $Net_{req}(d,f)$ (see eq. \ref{eq:device}).
\begin{equation}
    \label{eq:device}
    \forall d \in N, d = <Att(d), Cap(d), Net_{req}(d,f)>
\end{equation}
Each attribute $a \in Att(d)$ corresponds to a device-specific property that may derive users' preferences of using the device over the others. For example, a coffee maker can have several attributes such as \textit{brand, quality of coffee produced, time required}. On the other hand, the set of capabilities $Cap(d)$ of device $d$ refers to the group of functions that can be executed by the device. A device can have multiple attributes and can support one or multiple capabilities. For example, in Fig. \ref{fig:wf} the first coffee maker has attributes of \textit{Brand A} and can quickly prepare coffee (i.e. \textit{speed}), while it has only one capability which is making coffee. The network requirements $R$ for a device $d$ to execute a function $f \in Cap(d)$ is given by $Net_{req}(d,f)$ (see eq. \ref{eq:netreq}). Such that $r \in R$ is a tuple of required access to the network for a device $d$ to execute a function $f \in F$, where $F$ is the set of workflow functions. Network requirements $Net_{req}$ function will be utilised for automatic policy generation, to be introduced in Section \ref{sec:policygen}.

\begin{equation}
    \label{eq:netreq}
     \forall d \in D_s, f \in F: R = Net_{req}(f,d)
\end{equation}

Network requirements can be extracted based on several existing techniques such as capturing and analysing IoT traffic \cite{hamza2018clear,sorensen2017automatic}. In the context of this paper, the goal is to identify a list of protocols used by the device rather than the communication endpoints (e.g. IP address and port). Network requirements can also be extracted from MUD, which clearly defines the type of protocols used by specific devices. However, MUD can't specify precisely the endpoints that devices should have access to prior to deployment. However, we utilise the workflow to define the endpoints for a device.

A user drives the operations in a network by creating one or multiple activities as functional workflows for the network to fulfil. A functional workflow $W$ is represented as a directed acyclic graph (DAG) $W = <F,E>$, where $F$ is a set of functions, and $E$ is the dependencies among them. For a workflow $W$ to be feasible, we must ensure that each function $f \in F$ can be executed by at least one device in the network $N$ (see eq. \ref{eq:wfdev}). 

\begin{equation}
\label{eq:wfdev}
    \forall f \in F, \: \exists \: d \: ,  f \in Cap(d)
\end{equation}

Meanwhile every edge $ e \in E$ represents the dependencies between a pair of functions. For example, in Fig. \ref{fig:wf} the coffee making function depends on the alarm function. In other words, only after an alarm is triggered that a coffee maker can start to make coffee. To support this dependency, we should permit the communication between a specific alarm device and a specific coffee maker chosen to fulfil the workflow.


\section{User preference model}
\label{sec:prefmodel}
In order to determine the suitable devices for a workflow, we adopt a flexible \textit{user preference model} $M(F,D)$ to quantify user preferences of using any devices for various functions. \textit{user preference model} can answer questions like how likely a user will prefer to use any subset of devices $D \subseteq N$ for a given workflow of functions $F$. We can use any preference models flexibly in our proposed system. Such a users preference model can also be obtained by using a variety of machine learning techniques such as preference logic, fuzzy logic, neural networks \cite{pigozzi2016preferences}. 

\begin{figure}
    \centering
    \includegraphics[width=0.35\textwidth]{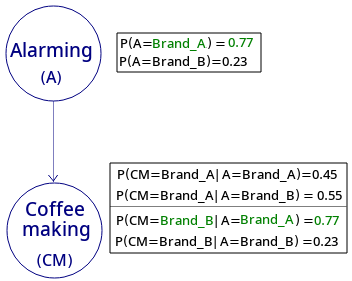}
    \caption{User preference representation using Bayesian network for the example in Fig. \ref{fig:wf}.}
    \label{fig:BN-alarm-cm}
\end{figure}

However in our experiments, we chose to use Bayesian networks technique to represent user preference model. Bayesian network \cite{daly2011learning} is a common technique for modelling user preferences and predict users activities \cite{wu2016bayesian,rim2013bayesian, rosenfeld2018study}. For example, the preference model in Fig. \ref{fig:BN-alarm-cm} can tell our system that our user prefers to use Brand\_A alarm device and Brand\_B coffee maker, their join probability is the highest among any other combination. According to the graphical structure of the Bayesian network, we can easily calculate the probability for the user to prefer using any group of devices together. Certainly, the higher the probability (i.e. the joint probability), the more preferable the corresponding group of devices would be to support any specific activities.


Bayesian network can be constructed based on historical data regarding devices usage in various workflows functions. In fact previously developed learning algorithms can learn the network structure (i.e. graph representation $DAG$) as well as the parameters (i.e. probability distribution $P$) from complete or incomplete data \cite{rim2013bayesian}. One approach is to find the network that maximises the likelihood of the data using search algorithms \cite{grossman2004learning}. In \cite{wu2016bayesian}, Wu et al. successfully construct Bayesian Network using activity dataset \cite{nazerfard2013using} to accurately predict user activity. Similarly Bayesian network can be constructed from a dataset of historical workflows. Moreover, users historical workflows include the dependencies between the functions, which can be used to guide the structure learning of the Bayesian network. User preference modelling is not the central focus of this paper. We rely on existing machine learning techniques and will not investigate this issue further.

\section{Problem formulation}
\label{sec:problem_formulation}
Given a network model $N$, user preference model $M(F,D)$, and a workflow $W = <F,E>$, how to determining the set of devices $D_s \subseteq N$. Such that $D_s$ fulfils all functions $F$ in the workflow $W$ and maximises user preference. 

We formulate this problem as following:

INPUT: $(N, W(F,E), M(F, D_i) )$. 

OUTPUT: $ d \in D_s \subseteq N $

OBJECTIVE:
\begin{equation}
\label{eq:problem}
    \argmax_{ D_{i} \subseteq N } M(F, D_i)
\end{equation}

s.t. $\forall f \in F \: \exists \: d \: ,  f \in Cap(d)$ 

Search algorithms will be used to find the set of devices that maximise user preference among all devices in the network $N$ that satisfy the flow requirements $F$. Such that for any set of devices $D_i$ a query will be sent to the preference model $M(D_i,F)$ to get the associated preference probability score. Then the devices with the highest score will be selected to implement the workflow functions.



\section{Solution searching methods}
\label{sec:searchingmethods}
We study three common optimisation algorithms to tackle the device selection problem in eq.(\ref{eq:problem}). We focus on hill climbing (HC) and simulated annealing (SA) as two representative local optimisation methods as well as genetic algorithm (GA) as a typical global optimisation method. We also study the Brut-Force search and to what extend it can scale. The intention is to test which algorithm is more suitable to solve this problem in terms of scalability, efficiency and solution quality. 

As a simple search technique, HC starts with arbitrary selection of devices and gradually improve users preferences by changing some of the devices that support the activity workflow. We choose to use Hill-climbing as it is highly efficient. However this technique can be easily trapped by local optima and may not scale well to large problems. To address this issue, we have also studied SA  \cite{Luke2013Metaheuristics} which can escape the local optima by allowing less optimal movement based on a probability function. GA is a well-known evolutionary computing algorithm and has been widely demonstrated to perform well on many difficult combinatorial optimisation problems. On the other hand, Brute-force algorithm is systematically enumerating all possible candidates for the solution and selecting the best candidate. It is guaranteed to find the optimal solution, however, it can't scale to large problem.

The pseudo code for device selection in  Algorithm \ref{alg:selection} requires a network $N$, a preference model $M$, and a workflow $W$ as inputs. Algorithm \ref{alg:selection} starts by filtering network devices and only select devices $D_{cands}$ that satisfy workflow requirements (i.e. $F$) (see eq. \ref{eq:dcand}). 

\begin{equation}
\label{eq:dcand}
    \forall d \in D_{cands} \subseteq N,  f \in F : \:   f \in Cap(d)
\end{equation}

The selected devices $D_s$ is initialised with a random device candidates. Then each of the searching algorithm performs the following steps until it reach its termination condition (based on the settings, to be discussed later); a) selects a new candidate solution $D_i=alg(D_{cands}, F)$; b) uses the preference model to score its accuracy $M(D_i,F)$; c) and decide whether to replace the existing candidate $D_s$ by $D_i$ based on its internal mechanism. 

\begin{algorithm}[ht]
 \KwData{$N, W<F,E>,M$}
 \KwResult{$score_s, D_s \subseteq N$}
  $D_{cands} =  flow\_req(N, F)$ \;
  $D_s = getRandomCand(D_{cands}, F)$ \;
  $score_s = M(D_s,F)$ \;
 \While{not termination condition}{
       $D_i =  search_{alg}(D_{cands}, F)$ \;
       $score_i = M(D_i,F)$ \;
       $score_s, D_s = search_{alg}(D_i,D_s)$  \;
       }
  \caption{Device selection searching algorithm}
  \label{alg:selection}
\end{algorithm}

\section{Policy generation}
\label{sec:policygen}
In the previous section, we discussed the algorithms to be exploited for addressing the device selection problem. In association with the solution to the problem a set of network access rules must be established to support the activity workflow and enforce the least privilege principle. We focus on network access control and assume authentication, such as key-based authentication \cite{ferrag2017authentication}, is supported in the network. Automatically generated policies can be subsequently enforced by any existing enforcing mechanisms developed for IoT such as Software Defined Networking \cite{bakker2016network}. 
 
The idea is to derive network ACLs policy for the workflow selected devices $D_s$ to fulfil the network requirements $R$ (see eq. \ref{eq:netreq}) for each device to execute the assigned function. 

Network requirements can include destination IP addresses, transport protocol, and port numbers, bandwidth, duration. In this work we only consider source and destination IP addresses, destination port, and the TCP protocol $\{src\_ip: \: source IP, dest\_ip:destination \: IP, dest\_port: destination \: port, tp\_proto: transport \: protocol\}$. These requirements can be easily mapped to ACL policy to guarantee that the selected devices $D_s$ can execute th workflow and obey the least privilege principle. The ACL policy can be easily enforced within and on the border of the network using software define networking \cite{bakker2016network}. 

\section{Evaluation}
\label{sec:eval}
We performed an empirical evaluation of four search algorithms to determine which one best suits our device selection problem (see eq. \ref{eq:problem}) for typical smart home IoT. The evaluation metrics are the quality of the solution in terms of user preference and time efficiency. 
\subsection{Experiment setup}
We fine-tuned the hyper-parameters for GA and SA and found that the parameter settings summarised in Table \ref{tab:GASAsettings} can enable the two algorithms to perform reasonably well.

 \begin{table}[t!]
        \caption{Experiment GP and SA parameters}
        \centering
        \begin{tabular}{| l | l |} 
            \hline
            \multicolumn{2}{|l|}{\textbf{Genetic Algorithm}} \\
            \hline
            \textbf{Parameter} & \textbf{Value} \\ 
            \hline
            Generations & 1000 \\ 
            \hline
            Population size & $|F|$ * 200 \\
            \hline
            Crossover rate & 0.7 \\
            \hline
            Mutation rate & 0.2 \\
            \hline
            Elitism rate & 0.1 \\  
            \hline
            Selection Method & Tournament \\  
            \hline
            Tournament size & 3 \\  
            \hline
            \multicolumn{2}{|l|}{ \textbf{Simulated Annealing}} \\ \hline
            \hline
            \textbf{Parameter} & \textbf{Value} \\ 
            \hline
            Steps & 200000 
            \\ \hline
            Max temperature & 0.2 
            \\  \hline
            Min temperature & 0.0001 
            \\  \hline            
        \end{tabular}
        \label{tab:GASAsettings}
    \end{table}

We compared algorithms in terms of scalability and efficiency by using multiple workflows composed of 4 to 7 functions (i.e. $|F| = 4, 5, 6, 7$), where each function has 7  alternative devices in the network $N$, see Table \ref{tab:netsettings}. Note the search space is $|F|^{7}$.

\begin{table}[ht]
\caption{Experiment Settings}
\label{tab:netsettings}
\centering
\begin{tabular}{|l|l|}
\hline
\textbf{No Workflow functions $|F|$} & 4, 5,6,7         \\ \hline
\textbf{Optimal user preference  $P(D_s)$} & 0.34, 0.42, 0.3, 0.24        
\\ \hline
\textbf{No Network Devices $|N|$}    & $rand(|F|, 4*|F|)$ \\ \hline
\textbf{No Device capabilities}    & $rand(2,7)$        \\ \hline
\textbf{No alternative devices}    & 7                \\ \hline
\textbf{Experiment runs}    & 30                
\\ \hline

\end{tabular}

\end{table}

A user preference model is constructed as a random Bayesian Network where each node represents a function and its values are the alternative devices. To verify the effectiveness of the search algorithms, we mimic user preference by tuning the Bayesian network probability distribution. We do that by randomly selecting a set of devices as preferred by the user and maximising their joint probability. This is achieved by simply by setting all conditional probabilities related to the desired collection of devices to be $p^{1/|F|}$ so that the joint probability will be $p$. For example, in Fig. \ref{fig:BN-alarm-cm} to set user preference to alarming using Brand\_A device and making coffee using Brand\_B we set their conditional probability to be $0.6^{1/2}$, the probability $P(CM=Brand\_B, A=Brand\_A) = 0.6$. Table \ref{tab:netsettings} shows the generated optimal user preference probability $P(D_s)$ to $(0.34, 0.42, 0.3, 0.24)$ for each composed workflow of $4,5,6 and 7$ functions respectively. These values are used in the evaluation to verify the search algorithms.

\subsection{Results and discussion}

\begin{figure}[ht]
    \centering
    \begin{subfigure}[b]{0.5\textwidth}
    \includegraphics[width=1\textwidth]{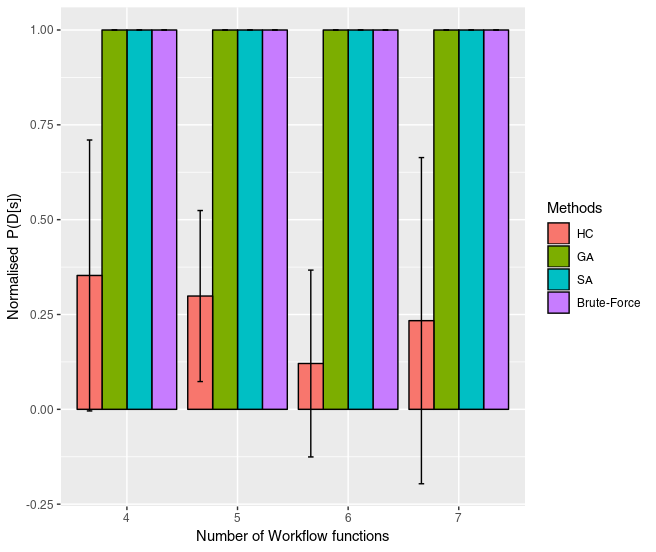}
    \caption{User preference optimisation}
    \label{fig:perf}
    \end{subfigure}
    \newline
    \begin{subfigure}[b]{0.5\textwidth}
    \includegraphics[width=1\textwidth]{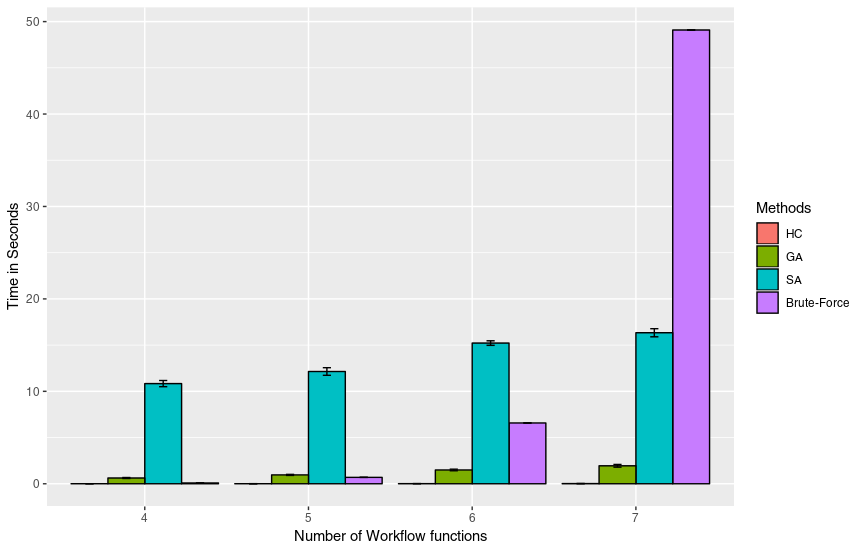}
    \caption{Time efficiency}
    \label{fig:time}
    \end{subfigure}
    \caption{Device selection heuristic algorithms comparison}
    \label{fig:score_time_results}
\end{figure}

The results shows that HC, as expected, is often trapped into local optima solutions, meaning it doesn't select devices for the workflow function that are most preferred according to the preference model. The effectiveness of HC deteriorates sharply with the increasing number of workflow functions, as shown in Fig. \ref{fig:perf}. SA and GA both can always find the optimal solution, see Fig. \ref{fig:perf}. We verify the optimality of the solution using the known user optimal preference probability shown in Table \ref{tab:netsettings}.

On the other hand, the search algorithms vary in terms of the time they take to reach the solution as shown Fig. \ref{fig:time}. HC is the most efficient algorithm in terms of elapsed time. However, its effectiveness is not satisfactory as it is not finding the optimal set of devices that maximise user preference probability, see Fig. \ref{fig:perf}. SA outperforms HC in terms of effectiveness, but at high cost of computation efficiency, since it clearly requires significantly longer computation time. However SA still better than Brute-Force search which shows an exponential increase in the time it takes with the increase of number of functions in a workflow. The best balance between effectiveness and efficiency can be achieved in our experiments by using GA. GA produces optimal solution like SA and Brute-Force and it scales better as shown in Fig \ref{fig:time}. Although Brute-Force is efficient in small problems, it fails to scale as GA. For example, in Fig. \ref{fig:perf} Brute-Force is more efficient than GA for problems with workflow of 4 or less functions, however, for workflow of 5 or more functions the time it takes increases exponentially.

\section{Conclusion}
\label{sec:conclusion}
Activity workflows are a common technique for specifying automation activities in smart environment. In existing systems, users have to specify the underlying devices to execute activity. We extend this by allowing users to specify abstract workflows that are instantiated for the particular environment. Previous research has not explored the automatic generation of concrete workflows and enforcement of least privilege based upon user requirements and preferences. We present an approach that decouples workflow requirements from the specific devices so devices can be selected on deployment to satisfy user preference. We also show how an automatic policy can be generated given network requirements for executing a particular workflow function. We formulate user activity automation as a constraint optimisation task and solve it using heuristic search algorithms. Our experiments show that for small network of IoT devices Brute-Force search is reasonably a good choice for optimising device selection. However, for larger network GA is the best, among the algorithms that we test, as it balance between efficiency and effectiveness.

\bibliographystyle{IEEEtran}

\end{document}